\documentclass[12pt,preprint]{aastex}

\usepackage{emulateapj5}
\def\hide#1{}

\shorttitle{Hard -to-Soft State Transition in Aql X-1}
\shortauthors{Yu \& Dolence}

\begin{document}

\title{A Hard-to-Soft State Transition during a Luminosity Decline of Aquila X-1}

\author{Wenfei Yu\altaffilmark{1,2} and Joshua Dolence\altaffilmark{3}}
\altaffiltext{1}{Shanghai Astronomical Observatory, 80 Nandan Road, Shanghai 200030, China; wenfei@shao.ac.cn}
\altaffiltext{2}{Center for Theoretical Astrophysics and Department of Physics, University of Illinois at Urbana-Champaign, 1110 West Green Street, Urbana, IL 61801.}
\altaffiltext{3}{Astronomy Department, University of Illinois at Urbana-Champaign, 1002 West Green Street, Urbana, IL 61801. }
\altaffiltext{4}{http://heasar.gsfc.nasa.gov/docs/xte/pca news.html}

\begin{abstract}
We have discovered a spectral transition from the low/hard (LH) state to the high/soft (HS) state when Aquila X-1 was {\em declining} in observations made with the {\it Rossi X-Ray Timing Explorer (RXTE)}. The 2--200 keV energy flux corresponding to the state transition is $1.1\times{10}^{-9}~ {\rm ergs~cm^{-2}~s^{-1}}$, an order of magnitude lower than observed in the past. The 2--200 keV peak flux of the following HS state is $1.6\times{10}^{-9} ~{\rm ergs~cm^{-2}~s^{-1}}$. The relation between the luminosity of the hard-to-soft state transition and the peak luminosity of the following HS state confirms the linear relation found previously. This implies that the luminosity of the hard-to-soft state transition is not determined solely by the mass accretion rate, but appears to be determined by the peak luminosity of the soft X-ray outburst. We also found that the LH-to-HS state transition occurred at a luminosity similar to that of the corresponding HS-to-LH state transition, i.e., there is no apparent hysteresis. These results provide additional evidence that the mass in the accretion disk affects the luminosity of the hard-to-soft state transition, and that the accretion flow that powers the LH state is related to the accretion flow that powers the HS state at a later time.  
\end{abstract}


\section{INTRODUCTION}
The low/hard (LH) state and the high/soft (HS) state are the two main X-ray spectral states identified in the Galactic black hole binaries (see the review by McClintock \& Remillard 2006 and references therein). Similar spectral states, namely the island state and the banana state (Hasinger \& van der Klis 1989), have been seen in the atoll sources in the neutron star low-mass X-ray binaries (LMXBs) (e.g. van der Klis 1994; Barret \& Vedrenne 1994). The two spectral states, as well as transitions between the two states, are also similar in terms of their X-ray energy spectra and variability properties (e.g., Yu et al. 2003). These observational properties suggest that the same physics is involved in both black holes and neutron stars during the state transitions.

Mass accretion rate has long been thought to determine the spectral state of an accreting black hole or neutron star; thus, the variation of mass accretion rate was thought to cause the transitions  between the states (e.g., Esin et al. 1997). However, recent observations show that mass accretion rate is not the only parameter that determines the spectral states (Homan et al. 2001; Smith et al. 2002; Yu et al. 2004,hereafter YKF04). Suggestions for possible other parameters include (1) coronal size (Homan et al. 2001), (2) the size of the accretion disk (Smith et al. 2002), (3) the past history of the location of the inner edge of the accretion disk (Zdziarski et al. 2004), and (4) the mass in the accretion disk (YKF04). 

The suggestion that the mass in the accretion disk affects state transitions is based on the correlation between the luminosity of the hard-to-soft state transition and the peak luminosity of the following HS state in the observations of two outbursts in each of the transients \mbox{Aql X-1}, \mbox{XTE J1550$-$564}, and a flaring neutron star LMXB \mbox{4U 1705$-$44} (YKF04). The correlation indicates that the higher the luminosity of the hard-to-soft state transition, the brighter the following HS state will be. Because a brighter outburst or flare normally has a duration longer than or similar to a dimmer outburst or flare, a brighter outburst is associated with the accretion of more mass onto the compact object. There is additional evidence that the hard, power-law spectral component is associated with the mass in the disk. In the black hole transient \mbox{GX 339$-$4}, the peak fluxes of the LH states in the seven outbursts seen by BATSE and HEXTE in the past 15 years are nearly linearly related to the outburst waiting time (Yu et al. 2007). The correlation also suggests that the earlier accretion flow that powers the LH state and the later accretion flow that powers the HS state are somehow related, suggesting that the outer disk plays a role in the generation of the hard spectral component that dominates the X-ray emission in the LH state (YKF04; Yu et al. 2007). A similar correlation in the flaring neutron star LMXB \mbox{4U 1705$-$44} (YKF04; W. Yu 2007, in preparation) indicates that the phenomena are associated with the variation of the mass accretion rate in the accretion flow, rather than the outburst mechanism of soft X-ray transients (e.g., Cannizzo et al. 1995).  

In this paper, we report our study of four outbursts of \mbox{Aql X-1} in which the flux peaks of the LH states during the outburst rise and the following HS states were covered by {\it RXTE} observations. It is especially noteworthy that we have identified a hard-to-soft state transition during a {\it luminosity decline} of the LH state in a very faint outburst in 2001, in which the following soft-to-hard state transition occurred at a similar luminosity level. This outburst is actually fainter than the 2005 outburst, which was a hard state outburst candidate, observed by {\it RXTE} and {\it INTEGRAL} (Rodriguez et al. 2006). We also show that these observations establish a nearly linear relation between the luminosity of the hard-to-soft state transition and the peak luminosity of the following HS state over a luminosity range of an order of magnitude. The results suggest that the hysteresis of state transitions (see Miyamoto et al. 1995) is not always present in the same LMXB. 

\section{OBSERVATIONS AND ANALYSIS}
All of the public {\it RXTE} observations of \mbox{Aql X-1} until 2006 June were used. As a  first step, we analyzed the {\it RXTE} standard products and the {\it RXTE} ASM (2--12 keV) light curve of \mbox{Aql X-1} following the method described in YKF04. We generated the hardness ratios between the HEXTE (15--250 keV) and the PCA (2--9 keV) count rates. The hardness ratios clearly show the LH and the HS states, corresponding to hardness ratios of about 0.3 and 0.02, respectively. The distinct hardness ratios help us identify the times corresponding to a hard-to-soft state transition as well as a soft-to-hard state transition in each outburst. 

The observations of \mbox{Aql X-1} around these state transitions were performed roughly once every two days. Thus, the time window corresponding to any hard-to-soft state transition is accurate to about two days. We were able to identify both hard-to-soft and soft-to-hard state transitions in 4 out of a total of 10 outbursts in the past 10 years. These transitions occurred on MJD 51,316, 51,820, 52,095, and 53,162, as identified by HEXTE (15--250 keV) to PCA (2--9 keV) hardness ratios dropping from 0.3 to 0.02. The corresponding observation IDs are 40047-01-01-00, 50049-01-04-03, 60054-02-02-04, and 90017-01-09-01. We marked the data points in these observations as squares in the 2 day averaged light curves of the ASM, the PCA (2--9 keV), and the HEXTE (15--250 keV) in Figure 1. Similarly, we identified the observations of the corresponding HS state peaks by selecting the peak PCA (2--9 keV) count rates after the state transitions. These HS state peaks occurred on MJD 51,324, 51,831, 52,103, and 53,167, and are marked as diamonds (Fig.~1). The corresponding observation IDs are 40047-02-03-01, 50049-02-03-01, 60054-02-04-00, and 90017-11-01-00. These measured luminosities approximately represent the actual peak luminosities of the HS states (see also the luminosity evolution obtained with the spectral fits presented by Maccarone \& Coppi 2003). We also identified a hard-to-soft state transition during the decay of the LH state in the faint outburst of \mbox{Aql X-1} around MJD 52,095. The luminosity decreased in the LH state before the hard-to-soft state transition and increased after the transition to the HS state. There is an earlier report of a similar phenomenon in the {\it persistent} black hole binary \mbox{GRS 1758$-$258} (Smith et al. 2001). 

\subsection{Spectral Analysis}

The goal of our spectral analysis is to estimate the X-ray fluxes of these peak states, as well as to derive the luminosity evolution during the small flare that occurred around MJD 52,100. We extracted PCA spectra in the 3--20 keV band and HEXTE cluster 2 spectra in the 20--200 keV band for the peaks of the HS and LH states, respectively.  For each observation, the longest time interval with continuous coverage of all  turned-on PCUs was chosen. The standard data analysis package FTOOLS (ver. 6.0.4) was used. We applied the {\it RXTE} PCA bright-source background model in the analysis of most of the observations. The faint-source background model was applied (according to the PCA background subtraction tutorial at the {\it RXTE} Guest Observer Facility Web site$^4$) in those observations when the PCA count rates fell below 40 counts s$^{-1}$ PCU$^{-1}$ after MJD 52,109 (see Fig.~2). Wherever a type I X-ray burst occurred in an observation, an interval of 100 s was  excluded in the extraction of the PCA energy spectra. The bursts contribute little in the extracted HEXTE spectra above 20 keV. Therefore, the HEXTE spectra were extracted for the entirety of the observation periods. 

Currently, there is no consensus on the spectral model of an accreting neutron star in an LMXB. The neutron star, the boundary layer, and the accretion disk may contribute to the X-ray flux. When source luminosity is low and the accretion disk recedes farther out, the X-ray emission from the neutron star may be a dominant component in the soft X-ray band, which would be like that of a blackbody. Therefore, each energy spectrum was fitted using XSPEC (ver. 12.2.1) with a model composed of photoelectric absorption (wabs), a single blackbody, a power law, and a  Gaussian line, which is fixed at 6.5 keV to account for the iron line. {\it Swift} XRT observations of Aql X-1 have shown that such a spectral model can fit the data (Wijnands et al. 2006; A. K. H. Kong 2006, private communication), suggesting that our approach to measuring X-ray flux is appropriate. The neutral hydrogen column density was fixed at the commonly used Galactic value along the line of sight ($N_{\rm H}=3.4 \times 10^{21} \rm~atoms~cm^{-2}$ for \mbox{Aql~X-1}; see Maccarone \& Coppi 2003).  By including systematic errors of 1\% or less in the spectra, each fit achieved a reduced $\chi^2$ less than 2, with the degree of freedom being 78. This indicates a reasonable agreement between the spectral shape of the simple model and the data, and meets our need to measure the source X-ray flux in the 2--200 keV band. In order to estimate the energy flux below 2 keV, we have calculated the bolometric flux of the single blackbody component for the study of the small flare shown in Figure 2 (see below). The analysis of some {\it Swift} XRT observations of Aql-1 has shown that a spectral  model composed of a single blackbody plus a power law, and a spectral model composed of a disk blackbody plus a power law can both fit the data; similar 0.5--2 keV fluxes were derived for both the blackbody component and the disk blackbody component (Wijnands et al. 2006). Therefore, the bolometric correction to the blackbody flux likely accounts for most of the energy flux below 2 keV, even if the soft spectral component is instead dominated by a disk blackbody component. 

\subsection{Results}
In Figure 2, we plot the 2--200 keV X-ray flux ({\it filled circles}) and the 2--200 keV X-ray flux excluding the blackbody component ({\it solid line}) obtained in the observations of the faint outburst during which \mbox{Aql X-1} started a hard-to-soft state transition, around MJD 52,095. The source first rose to the peak flux of the LH state around MJD 52,085, then declined in the LH state for about 10 days. When the source was almost undetectable by the  ASM on MJD 52,095, the source started a transition to the HS state and rose in luminosity to its peak flux in the HS state around MJD 52,103. Then the source transited from the HS state to the LH state at around MJD 52,105. We can calculate the corresponding luminosities from the flux measurements. The conversion factor from the 2--200 keV energy flux (in units of $10^{-10}~{\rm ergs~s ^{-1}~cm^{-2}}$) to luminosity (in units of $10^{35}~{\rm ergs~s^{-1}}$) is 0.75 for a source distance of 2.5 kpc (Chevalier et al. 1999), or 3.0 for a source distance of 5 kpc (Rutledge et al. 2002), assuming the X-ray emission is isotropic. As shown in the upper panel of Figure 2 ({\it shaded regions}), the hard-to-soft and the soft-to-hard transition fluxes overlap and are similar in range to within a factor of less than 2; therefore, no apparent hysteresis of state transitions is seen. The transition luminosities are in the range $5.3-11.3 \times{10}^{35}~{\rm ergs~s^{-1}}$ (for a distance of 2.5 kpc) or $2.1-4.5\times{10}^{36}~{\rm ergs~s^{-1}}$ (for a distance of 5 kpc). For a neutron star of 1.4 $M_{\rm \odot}$, this corresponds to 0.5\% of the Eddington luminosity (for 2.5 kpc) or 2\% of the Eddington luminosity (for 5 kpc), respectively. 

We have also shown the evolution of the blackbody temperature and the corresponding blackbody bolometric fluxes in Figure 2. It is worth noting that the blackbody temperatures were mostly around 1.4 keV. Therefore, there is not much energy flux from the blackbody component below the PCA low-energy boundary of 2 keV. This implies that the decrease of the 2--200 keV flux around MJD 52,090, during which a hard-to-soft state transition occurred, indeed corresponds to a luminosity decline. 

There is a nearly linear relation between the HEXTE count rate corresponding to the hard-to-soft state transition and the ASM peak count rates. A linear fit gives a reduced $\chi^2$ of about 1100, suggesting that the data scatter significantly around the linear model.  If we include the effect of data scattering around the model, we get $C_{\rm HEXTE,P}=(0.7\pm5.7)+(2.3\pm0.3)C_{\rm ASM,P}$, where $C_{\rm HEXTE,P}$ and $C_ {\rm ASM,P}$ are HEXTE peak rates and ASM peak rates, respectively. This empirical relation may be used to predict the luminosity of a future outburst using hard X-ray observations during the outburst rise. We also plot the relation between the corresponding 2--200 keV fluxes in Figure 3. As expected, this relation is nearly linear as well. The linear Pearson correlation coefficient is 0.999. A linear fit gives a large reduced $\chi^2$, suggesting that the data scatter around the linear model as well.  If we include the scattering effect, we get $F_{\rm HS,P}=(-0.372\pm0.153)+(1.720\pm0.026)F_{\rm ST}$, where $F_{\rm HS,P}$ is the peak flux of the HS state and $F_{\rm ST}$ is the flux corresponding to the hard-to-soft state transition. The relation is tight, suggesting a that prediction of an outburst peak flux is possible when the hard-to-soft state transition is observed. Since the energy spectra among the LH states and among the HS states of different outbursts of the same source are nearly identical (e.g., Yu et al. 2003), the nearly linear relation between the energy fluxes suggests a similar relation between the luminosity of the hard-to-soft state transition and the peak luminosity of the following HS state of Aql X-1.  

\section{DISCUSSION AND CONCLUSIONS}
We have studied four outbursts of \mbox{Aql X-1} in which the hard-to-soft state transitions were covered by {\it RXTE} observations. In one of the four outbursts, we have discovered a hard-to-soft state transition which occurred during a luminosity decline and at a luminosity similar to that of the following soft-to-hard state transition. We have also found that the peak luminosities of the HS states follow the X-ray luminosities of the hard-to-soft state transitions nearly linearly. The study confirms the luminosity correlation previously found by Yu et al. (YKF04) and extends it to much lower luminosities. Using the best-fit linear relation, we can predict the peak luminosity of an outburst in \mbox{Aql X-1} at the very beginning of the hard-to-soft state transition. 

The observation of the hard-to-soft state transition during a luminosity decline demonstrates that the hard-to-soft state transition is not associated with the flux peak of the LH state before the state transition. In all other outbursts of \mbox{Aql X-1}, as well as in the outbursts of several other transient sources such as \mbox{GX 339$-$4} (Yu et al. 2007), the state transitions were associated with the luminosity peaks of the LH states (Yu et al. 2003, 2007; YKF04). Thus, it is the luminosity of the hard-to-soft state transition, instead of the peak luminosity of the LH state before the state transition, that is correlated with the peak luminosity of the following HS state. 

In the most popular picture of spectral states in X-ray binaries, e.g., the model proposed by Esin et al. (1997), spectral states are determined by the mass accretion rate. A source in the LH state can only transit to the HS state when the mass accretion rate increases. Therefore, the hard-to-soft state transition only occurs during a luminosity increase. The observation of a hard-to-soft state transition during a luminosity decrease in \mbox{Aql X-1} contradicts this idea, implying that other parameters in addition to the mass accretion rate also determine the spectral transition. Since the luminosity of the hard-to-soft state transition can vary by an order of magnitude and that of the soft-to-hard state transition is usually at a much lower luminosity and remains nearly constant (Maccarone 2003), the hysteresis of the state transitions (e.g., Miyamoto et al. 1995) likely originates from the large variation of the luminosity of the hard-to-soft state transition. Therefore, the physics of the hard-to-soft state transition should be the focus.

The nearly linear relation between the luminosity of the hard-to-soft state transition and the peak luminosity of the following HS state indicates that the luminosity at which the hard-to-soft state transition occurs appears correlated with the peak luminosity of the following HS state, and is thus related to the properties of the accretion flow which powers the soft X-ray outburst. This observation challenges the interpretation that the variation in the luminosity of state transition is due to different coronal sizes (e.g. Homan et al. 2001). Since there is a tight correlation between the luminosity of the hard-to-soft state transition and the peak luminosity of the following HS state, the cause of the variation in the luminosity of the hard-to-soft state transition is also the cause of the variation of the peak luminosity in the HS state. In the HS state, the thermal spectral component dominates, and neither a static spherical corona nor a static disk corona could power the source or collapse to form a disk flow to power the source in the HS state. Therefore, coronal properties, such as coronal size or height, can not be the additional that parameter determines the spectral state transitions. The observation is also inconsistent with the interpretation invoking the history of the location of the inner-disk radius (Zdziarski et al. 2004). According to this interpretation, the smaller the initial inner disk radius, the faster a transient source reaches a hard-to-soft state transition, and thus yields a lower transition luminosity. There is no correlation between the luminosity of the hard-to-soft state transition and the peak luminosity of the HS state. 

The idea that disk size affects state transitions originated from the proposed geometry of two independent flows, which is based on the long-term correlation between power-law indices and fluxes (Smith, Heindl \& Swank 2002). According to this interpretation, the mass accretion rate in the disk flow is suspected to be an integration of that of a halo flow in the recent past which powers the LH state, because there is a {\it significant} difference between the viscous timescales in the disk flow and the halo flow. 

Although our suggestion is similar to the two-flow idea, there are significant differences. Our suggestion is based on the observations of individual state transitions and the correlation between the fluxes, which is shown in Figure 3. If the two-flow geometry is correct during the state transitions, the two flows have to be related in instantaneous mass accretion rate, instead of being related in a way involving integration of mass accretion rate over time (see also YFK04). A reduced integration effect is also suggested by the timescale of the rise of the LH state and the rise of the HS state during the start of an outburst, which differ by only a factor of 2. Therefore, the halo flow, if there is any, should be a sub-Keplerian flow which is very close to a Keplerian flow (see also YFK04). It is worth noting that instead of a halo inflow, we also regard an outflow as a possible alternative that would power the source in the LH state. 

It is also known that \mbox{Aql X-1} and other sources show ``parallel tracks" in the color-count rate or QPO frequency-count rate plot (e.g., Mend\'ez et al. 1998) which would be explained with a two-flow geometry (van der Klis 2003). The hard-to-soft state transitions occurring at different luminosities are actually extreme cases of the parallel tracks. Therefore, the mechanism for the variation in  the state transition luminosity is likely the same one that causes the parallel tracks. 

Our current analysis determines that it is the luminosity of the hard-to-soft state transition, instead of the peak luminosity of the LH state, that is correlated with the peak luminosity of the soft X-ray outburst. Therefore, there is only an underlying positive correlation between the instantaneous mass accretion rate at the transition and that at the luminosity peak of the HS state. The relation suggests that the X-ray radiation in the LH state is related to the optically thick disk flow further out, which, at a later time, would approach the innermost region and power a HS state. Because of the positive correlation between the luminosities, the optically thick disk in the outer region is probably the ultimate source which powers the hard spectral component originating from the Comptonization of hot electrons. Our study suggests that the disk flow may split part of its mass flow to contribute to a halo flow or an outflow on its way towards the compact star. Thus, at a given time, the flux of the LH state is an integration of the contribution from the disk over a range of radii which would vary in different outbursts. We speculate that the competition between the inner disk, which radiates soft photons and cools hot electrons, and the outer disk, which probably heats hot electrons, determines the spectral transition through the process of Comptonization. Assuming the mass accretion rate is a function of time and radius, the hard-to-soft state transition probably occurs when the inner disk dominates as a result of the mass accretion rate at the outer disk declining faster than that of the inner disk or increasing slower than that of the inner disk.  The soft-to-hard state transition occurs when the outer disk dominates as a result of the mass accretion rate through the inner disk declining faster than that of the outer disk or increasing  slower than that of the outer disk. This speculation is consistent with the idea that the mass in the disk somehow determines the state transitions, since at the beginning or at the end of an outburst, most of the mass is in the disk that is farther out. However, it is worth noting that only when the disk mass accretion rate reaches a certain threshold can the disk flow survive diminishment due to a halo flow or outflow, reach the inner most region, and form a HS spectral state. 

\acknowledgments
We thank the anonymous referee for a careful reading of the manuscript and useful comments, which improves the paper significantly. We would like to thank Rob Fender of the University of Southampton for useful discussions. This work was partially supported by NASA grants NNG 05GL60G,  NAG 5-12030 and NAG 5-8740, and NSF grant AST 0098399 at the University of Illinois, and by the starting funds at SHAO associated with the One Hundred Talents Project of the Chinese Academy of Sciences. The work has made use of data obtained through the High Energy Astrophysics Science Archive Research Center Online Service, provided by the NASA/Goddard Space Flight Center.

\clearpage

\begin{figure}
\includegraphics[scale=.80,angle=90]{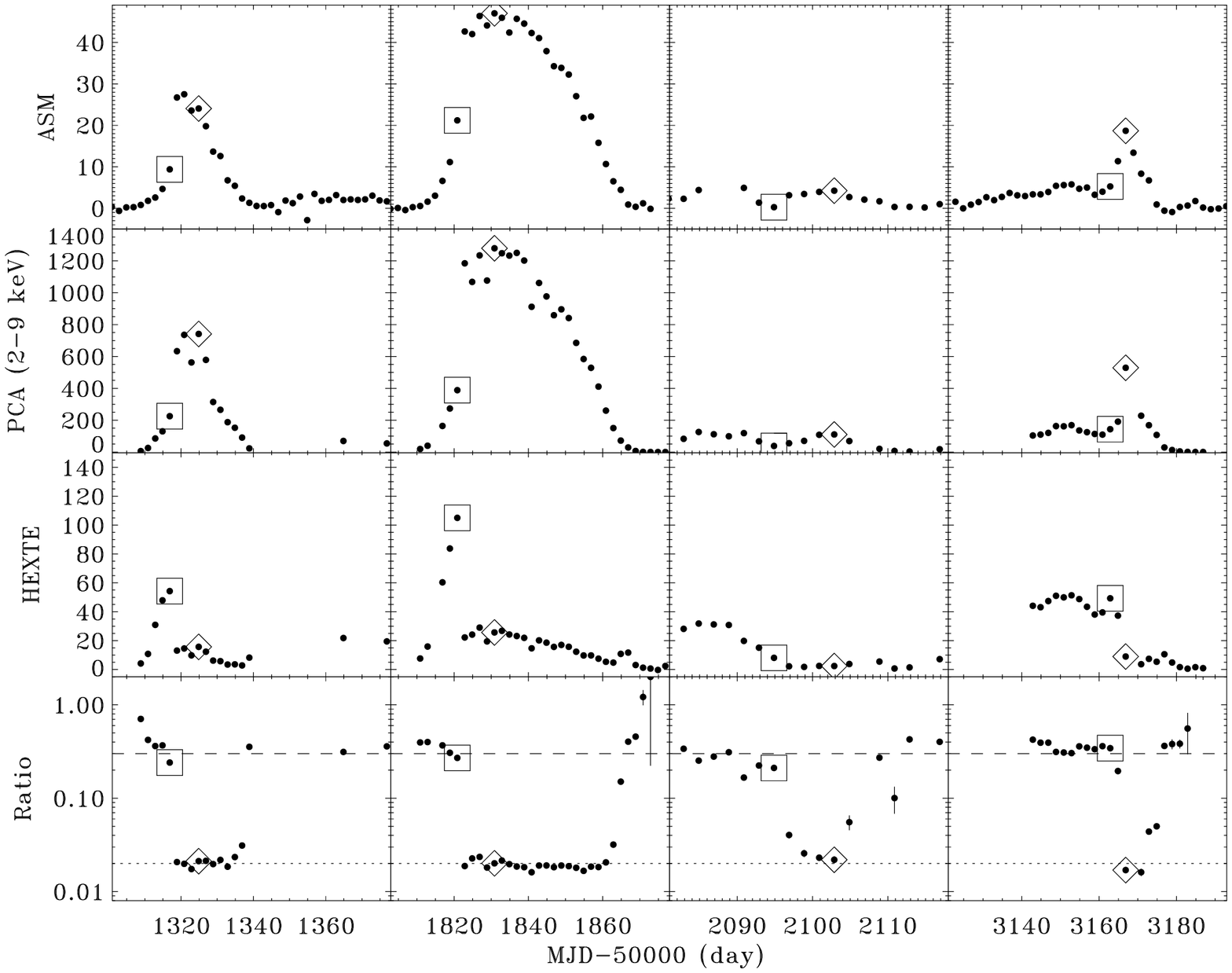}
\caption{Identification of hard-to-soft state transitions and luminosity peaks of the four outbursts of \mbox{Aql X-1} in {\it RXTE} observations. From top to bottom, the figure shows the ASM (2--12 keV) daily averaged count rates, the PCA (2--9 keV) count rates, the HEXTE (15--250 keV) count rates, and the count rate ratios between the HEXTE (15--250 keV) rates and the PCA (2--9 keV) rates for the four outbursts during which the hard-to-soft state transition was covered by {\it RXTE} pointed observations. The third column shows the outburst in which the hard-to-soft state transition occurred during a luminosity decline. Error bars are drawn, but are not visible because they are smaller than the symbol size. }
\end{figure}

\clearpage

\begin{figure}
\epsscale{.80}
\plotone{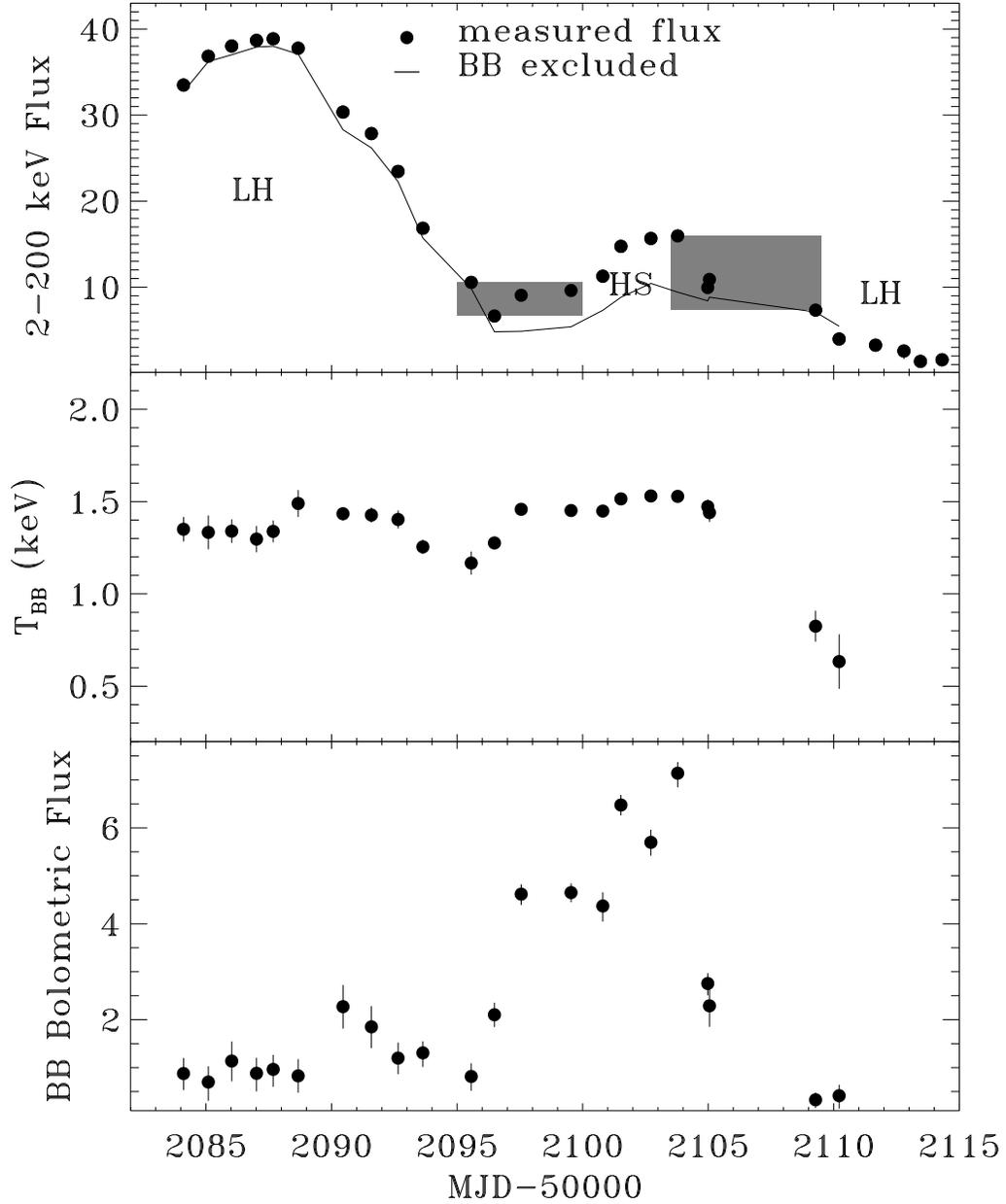}
\caption{Outburst of \mbox{Aql X-1} around MJD 52,095. {\it Top}: Evolution of the source flux (2--200 keV, in units of $10^{-10}~{\rm ergs~s^{-1}~cm^{-2}}$, {\it filled circles}) during the outburst in which a hard-to-soft state transition occurred during a luminosity decline. Error bars are drawn but are mostly not visible. The energy flux excluding the blackbody spectral component is shown as a solid line. The shaded regions show the state transition periods, i.e., from the start to the end of the hard-to-soft state transition and from the start to the end of the soft-to-hard state transition, respectively. There is no apparent hysteresis of state transitions. {\it Middle}: Evolution of the temperature of the blackbody spectral component as measured with the {\it RXTE} PCA. {\it Bottom}: Bolometric flux of the blackbody spectral component (in units of $10^{-10}~{\rm ergs~s^{-1}~cm^{-2}}$). The HS state was associated with an enhanced blackbody flux.  }
\end{figure}

\clearpage

\begin{figure}
\epsscale{1.0}
\plotone{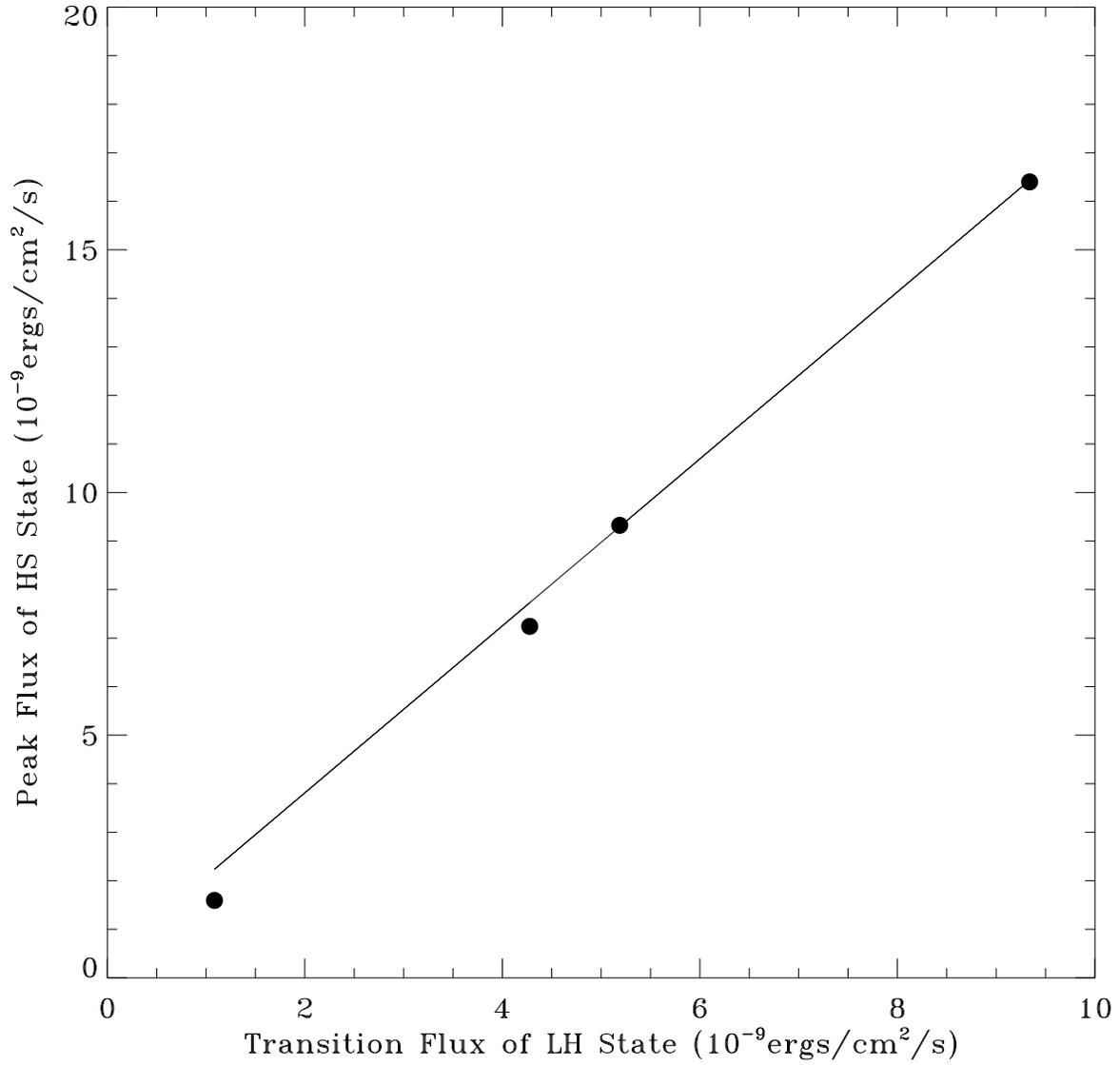}
\caption{Relation between 2--200 keV peak flux of HS state and 2--200 keV flux of LH state corresponding to start of hard-to-soft state transition. Error bars are drawn but are not visible. The best-fit linear relation is plotted as a straight line. }
\end{figure}

\end{document}